\documentclass{PoS}

\title{Precise predictions for charged Higgs boson production}

\ShortTitle{}

\author{\speaker{Maria Ubiali}\thanks{Funded by the Royal Society as a
  Royal Society Dorothy Hodgkin Research Fellow}\\
        Cavendish Laboratory, J.J. Thomson Ave,
Cambridge, CB3 0HE, United Kingdom\\
        E-mail: \email{ubiali@hep.phy.cam.ac.uk}}

\abstract{We review some of the major progress in the accurate calculation of the cross section for the production of a charged Higgs boson in a generic two-Higgs double model, by focussing on the high-mass region, for which new calculations have been made available in the past few years, both at the level of total and of differential cross sections. Furthermore, we illustrate some recent progress in the calculation of the total cross section for a charged Higgs boson with mass in the intermediate range $m_{H^\pm}\sim m_t$, including top-resonant contributions.}

\FullConference{Prospects for Charged Higgs Discovery at Colliders\\
		3-6 October 2016\\
		Uppsala, Sweden}

\newcommand{\hwbb}{pp\to H^{\pm} W^{\mp} b \bar b}
\newcommand{\ch}{H^{\pm}}

\begin{document}

\section{Introduction}
The detection of a charged Higgs boson would inexorably point to a broader Higgs sector than the one predicted by the Standard Model (SM). Thus, its discovery would be an unmistakable sign of the presence of new physics. For this reason 
extensive searches have been carried out by the ATLAS and CMS collaborations at the LHC. The analyses performed at the centre-of-mass energy of 7 TeV~\cite{Aad:2012tj,Aad:2012rjx,Aad:2013hla,Chatrchyan:2012vca}, 8 TeV~\cite{Khachatryan:2015uua,Khachatryan:2015qxa,Aad:2015typ,Aad:2014kga} and more recently 13 TeV~\cite{Aaboud:2016dig,ATLAS:2016qiq,CMS:2016szv} set stringent limits on the parameter space of the models featuring the presence of charged scalars. Experimental searches so far focussed on the detection of a light charged Higgs, with mass below the top quark mass (typically $m_{\ch}<160$ GeV),  or of a heavy charged Higgs, with mass above the top quark mass (typically $m_{\ch}>200$ GeV). In the light-mass range the main charged Higgs production mechanism is via the production of top-antitop pairs, with the (anti)top quark decaying into a positively(negatively) charged Higgs and (anti)bottom quark. 
This region is suitably described by the top-antitop production cross section multiplied with the branching ratio of the top decay, see for example Refs.~\cite{Czarnecki:1992zm, Denner:1990ns, Li:1990qf, Blokland:2005vq, Brucherseifer:2013iv}.
In the heavy-mass region, the charged Higgs bosons are mostly produced in association with top (anti)quark, as we discuss in more details in Section~\ref{sec:heavy}. In the intermediate region both mechanisms contribute and their interference must be consistently taken into account, as it is illustrated in Section~\ref{sec:intermediate}. 
\begin{figure}[b]
    \centering
    \includegraphics[width=0.35\textwidth]{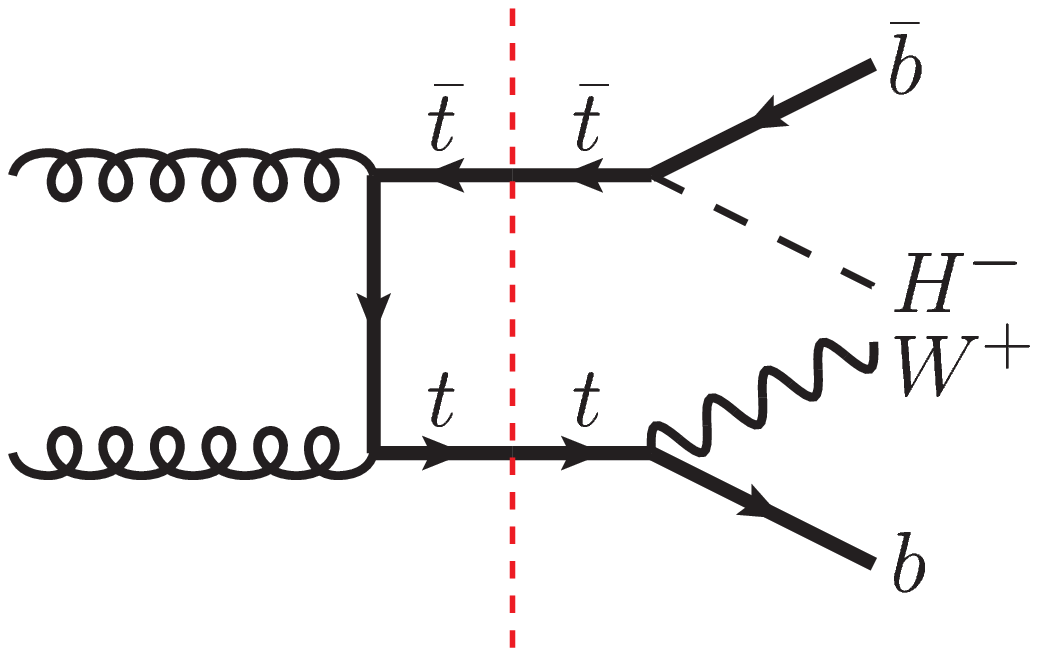} 
    \includegraphics[width=0.3\textwidth]{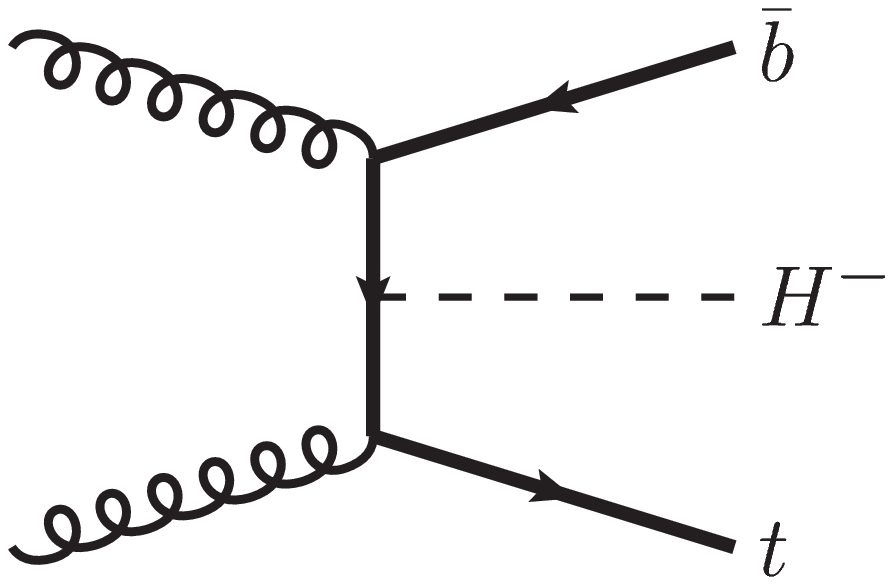}  
    \caption{\label{fig:diaglowhighmass} Sample tree-level diagrams for light (left) and heavy charged Higgs boson production (right).}
\end{figure}
In this contribution, we first summarise the main recent developments in the precise calculation of the total and differential cross sections for the production of a heavy charged Higgs boson. We then review recent progress in the precise computation of the inclusive cross section in the intermediate-mass range.
Although charged Higgs bosons appear in several BSM scenarios, here we focus on the simplest extension of the Standard Model, namely on a two-Higgs doublet model (2HDM), in which two isospin doublets are introduced
to break the $SU(2)\times U(1)$ symmetry, leading to the existence of five physical Higgs 
bosons, two of which are charged particles ($H^{\pm}$). 
Imposing flavour conservation, there are four possible ways 
to couple the SM fermions to the two Higgs doublets~\cite{Branco:2011iw}. Each of these four ways
gives rise to rather different phenomenological scenarios. The results reviewed in this contribution are displayed for a type-II scenario, but are easily generalisable to all other scenarios, by suitably rescaling the couplings of the charged Higgs with bottom and top quarks.

\section{Heavy charged Higgs boson production}
\label{sec:heavy}
\begin{figure}
     \includegraphics[width=.5\textwidth]{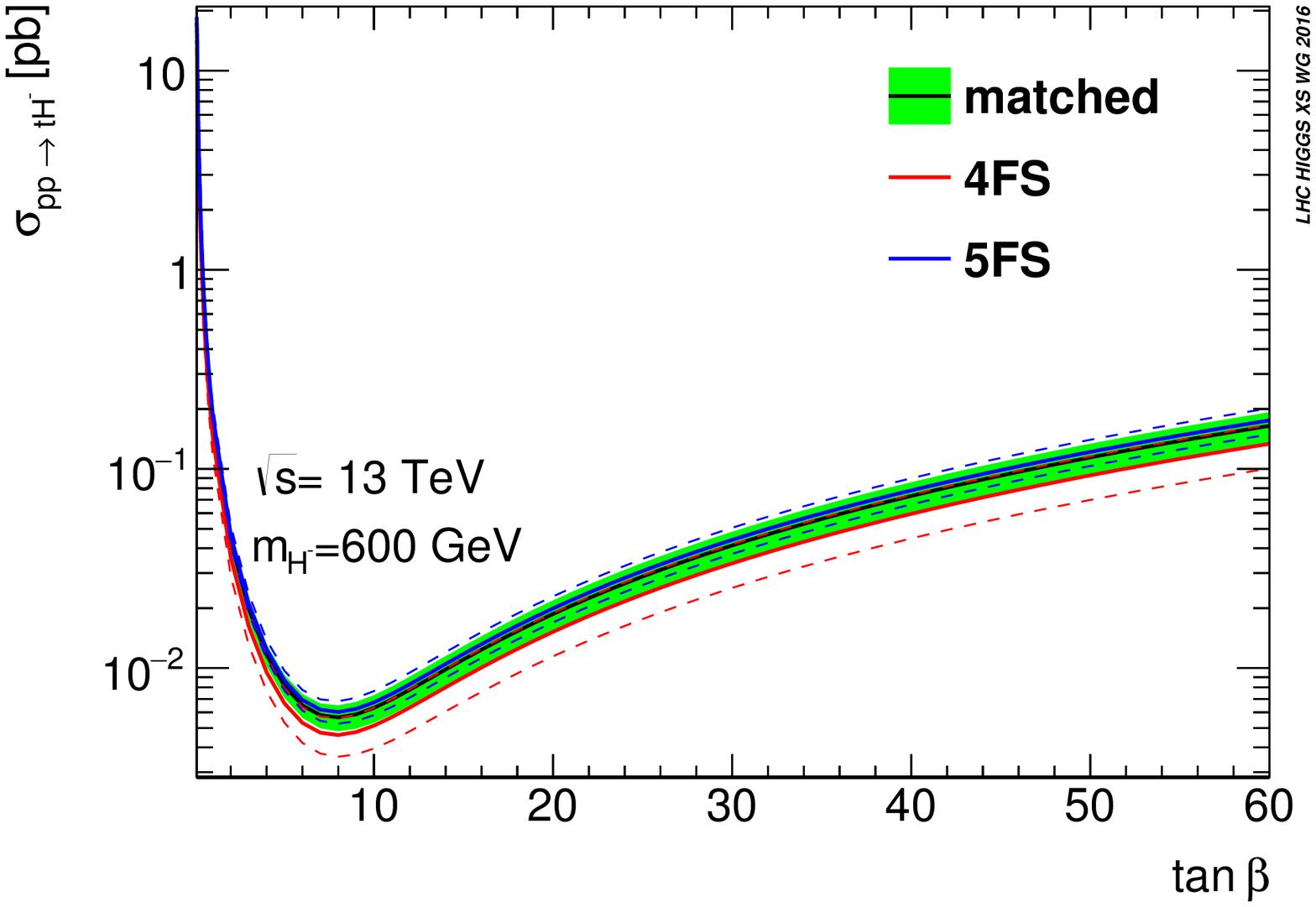}
     \includegraphics[width=.5\textwidth]{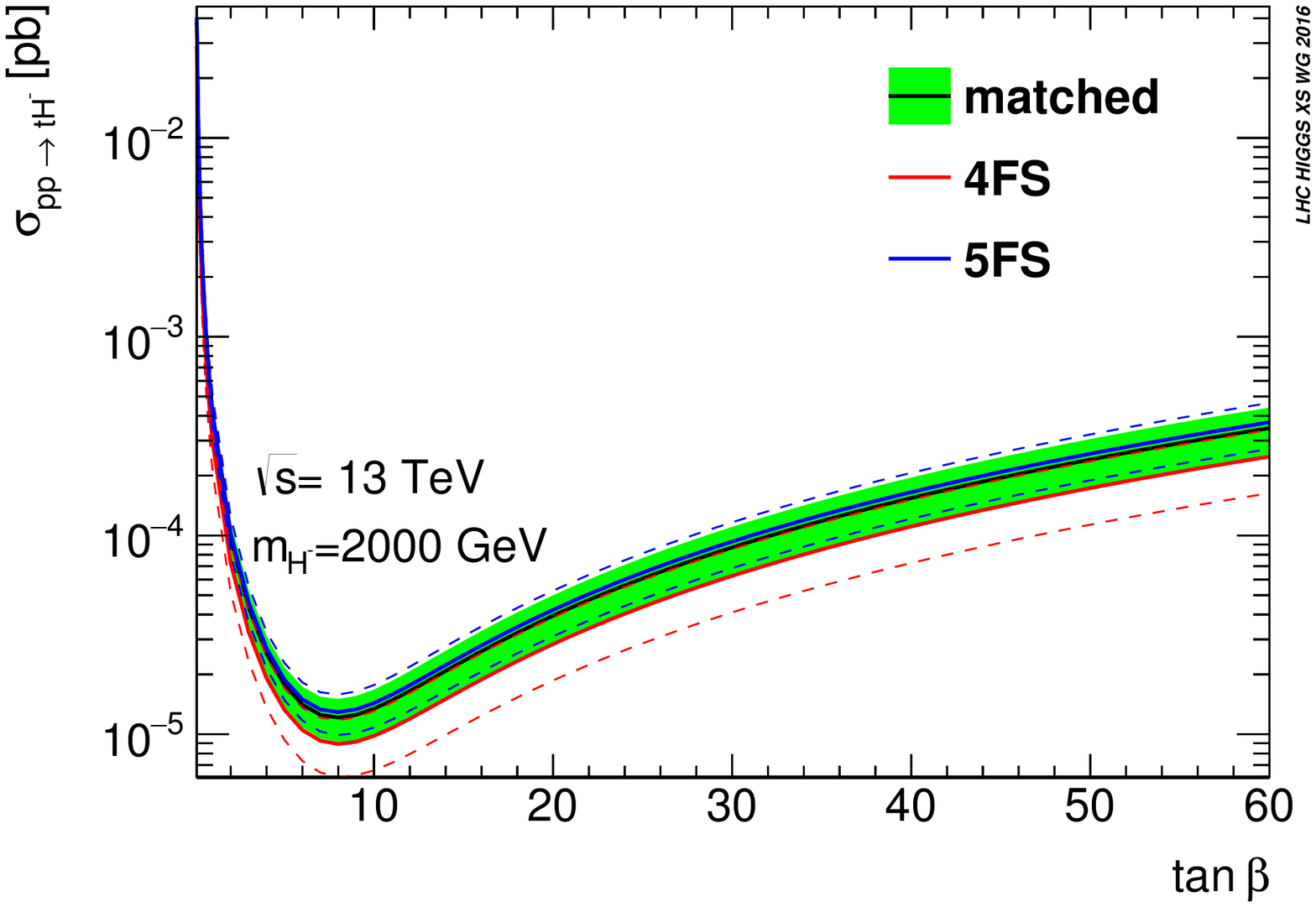}
\caption{\label{fig:totalheavy} Cross section for $tH^- + X$ production, after matching the 4FS (red) and 5FS (blue) results. The result is given for different values of $\tan\beta$ in the rang (0.1,...,60) for $m_{\ch}=600$ GeV (left) and 2000 GeV (right) at the LHC 13 TeV. The uncertainty band includes scale uncertainty associated with missing higher orders in the computation, PDF$+\alpha_s$ uncertainty, $m_b$ and $\mu_b$ uncertainties, $m_b$ being the pole mass of the bottom quark in the 4FS calculation and $\mu_b$ being the threshold at which the bottom PDF is dynamically generated off gluons and quarks in the 5FS.}
\end{figure}

Heavy charged Higgs bosons have mass larger than the top-quark mass and are dominantly produced in association with a top quark. Such a process, featuring bottom quarks in the initial state, can 
be described in QCD either in a four-flavor (4FS) or five-flavor scheme (5FS). In the former, 
the bottom quark mass is considered on the same footings as the other hard scales of the process 
and bottom quarks do not contribute to the proton wave-function. They can only be generated as massive final states at the level of the short-distance cross section. A representative tree-level Feynman diagram in the 4FS is depicted in the right panel of Fig.~\ref{fig:diaglowhighmass}. 
Instead, in five-flavour scheme, the bottom quark mass is considered 
to be a much smaller scale than the hard scales involved in the process and 
bottom quarks are treated on the same footing as all other massless partons, thus the tree-level diagram is initiated by a bottom-quark.
Next-to-leading order (NLO) calculations for the total cross sections in the 5FS~\cite{Zhu:2001nt,Gao:2002is,Plehn:2002vy,Berger:2003sm} including super-symmetric corrections have been performed more than fifteen years ago. 
Threshold resummation effects have also be computed up to NNLL accuracy~\cite{Kidonakis:2016eeu}.
The NLO calculation for the total cross sections in the four-flavor scheme is more involved, featuring an additional massive final state in the leading-order matrix element. It was carried out about ten years ago~\cite{Peng:2006wv,Dittmaier:2009np} and it consistently includes electro-weak and super-symmetric corrections. 
In Ref.~\cite{Flechl:2014wfa} an up-to-date comparison of the next-to-leading-order total cross section in the 4FS and 5FS was presented. A motivated choice of factorisation scale, $\mu=(m_t+m_{\ch})/k$ with $k$ in the range 4-6, motivated by the study in Ref.~\cite{Maltoni:2012pa} brings the two predictions much closer to each other and reconcile them within the estimated theoretical uncertainties due to missing higher-order corrections, parton distribution functions and physical input parameters. A four- and five-flavour scheme matched prediction using the Santander Matching weighted average~\cite{Harlander:2011aa} is provided for the interpretation of current and future experimental searches
for heavy charged Higgs bosons at the LHC. The predictions have been recently updated by using the most recent PDF sets and parameter settings in the Higgs Cross Section Working Group Yellow Report 4~\cite{deFlorian:2016spz}. The results of the most recent comparison and matching are displayed in Fig.~\ref{fig:totalheavy}.\\
The four- and five-flavor scheme
computations of the charged Higgs production cross sections could be consistently matched, as it has been recently done in the case of bottom-fusion-initiated Higgs production in Refs.~\cite{Forte:2015hba,Forte:2016sja}. 
Instead of an interpolation between the four and five-flavor scheme results by mean of a weighted average, as in the case of the Santander matching, one could use a more systematic approach which preserves the perturbative accuracy of both computations by expanding out the 5FS computation in powers of the strong coupling $\alpha_s$, and replacing the terms which also appear in the 4FS computation with their massive-scheme counterparts. The result would then retain the accuracy of both 4FS and 5FS: at the massive level, the fixed-order accuracy corresponding to the NLO, and at the massless level, the logarithmic accuracy of the starting five-flavor scheme computation (NLL). This is a direction to be pursued in future studies.

As far as predictions at the fully differential level including parton shower (PS) effects are concerned, they were made available a few years ago in the POWHEG~\cite{Klasen:2012wq} and MC@NLO formalisms~\cite{Weydert:2009vr}. Both computations were performed in the five-flavor scheme.
\begin{figure}[ht]
\begin{center}
     \includegraphics[width=.4\textwidth]{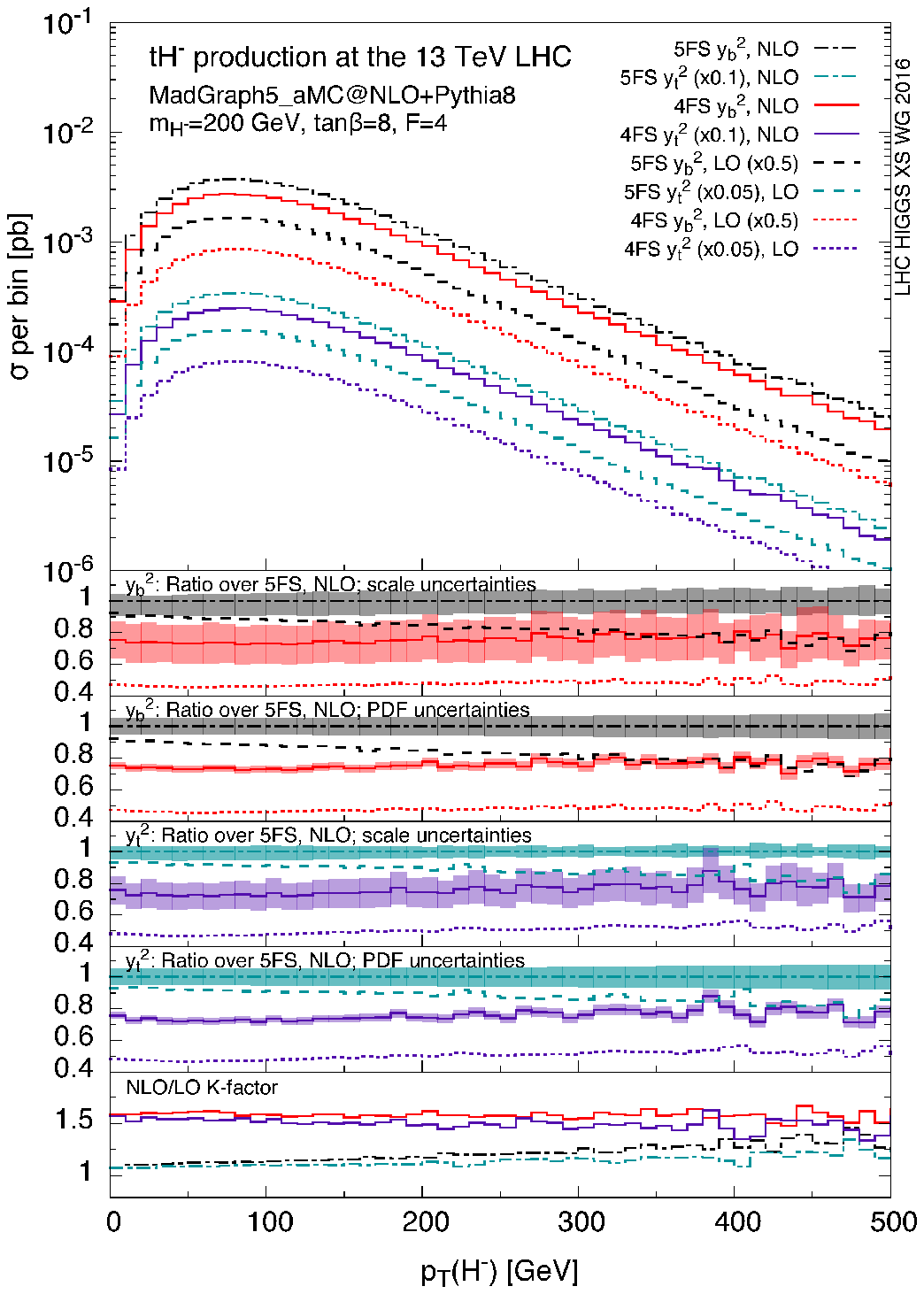}
     \includegraphics[width=.4\textwidth]{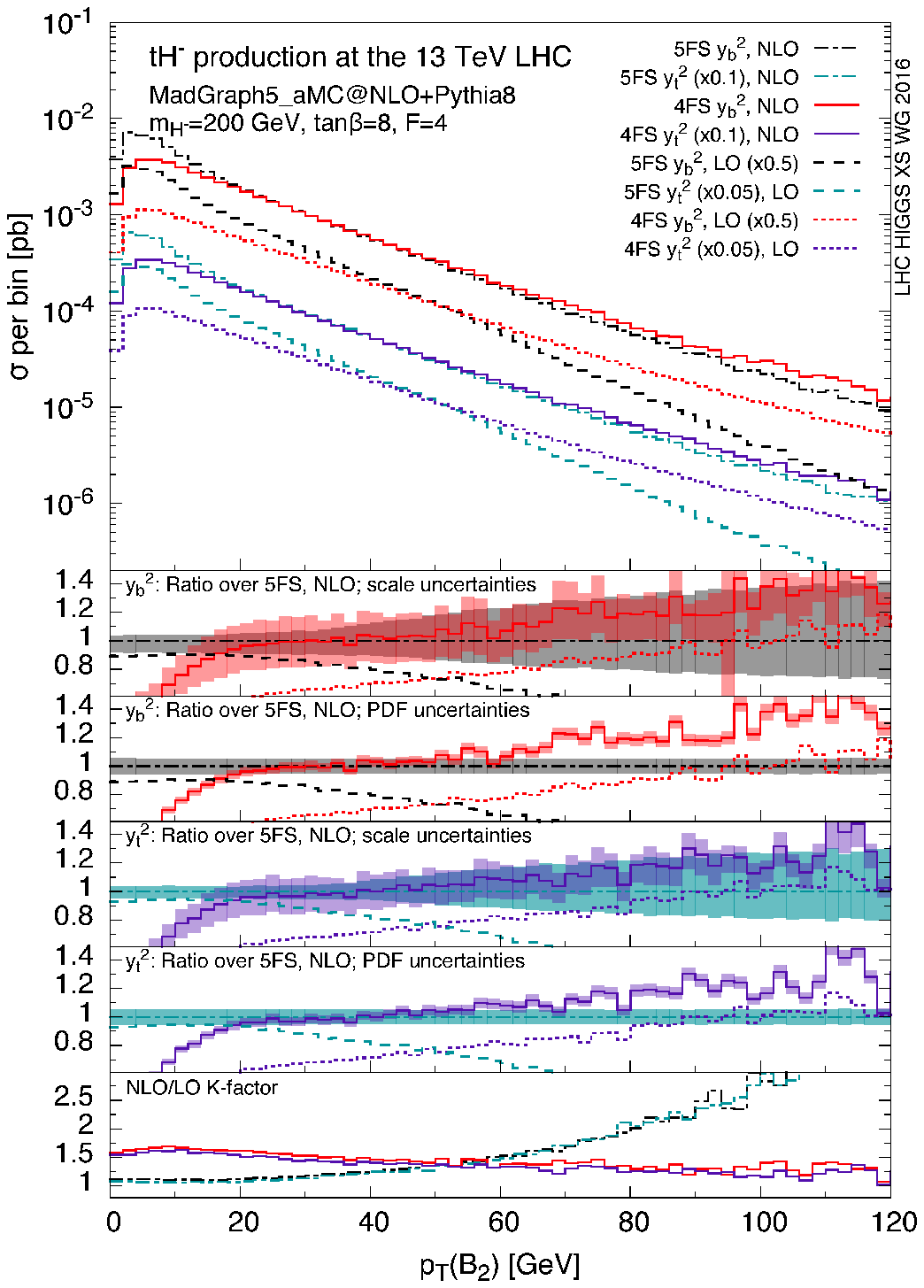}
\caption{\label{fig:diffheavy} LO and NLO predictions matched with {\tt Pythia8}~\cite{Sjostrand:2007gs} in the 4FS and 5FS, separately for the terms proportional to Yukawa coupling of the charged Higgs with the bottom quark ($y_b^2$) and the top quark ($y_t^2$). Left panel: transverse momentum of the charged Higgs boson. Right panel:  transverse momentum distribution of the second hardest $B$ hadron. Rescaling factors are introduced in the main
frame for better visibility. The first four smaller frames at the bottom show the ratio over the NLO prediction in
the 5FS for the $y_b^2$ and $y_t^2$ terms, and the scale and PDF uncertainty bands for the NLO curves.
The bottom frame shows the differential $K$ factor (NLO/LO) for the four predictions. A charged Higgs boson mass of $m_{\ch} = 200$ GeV is assumed.}
\end{center}
\end{figure}
Differential calculations were made available for the first time in the four-flavor scheme in Ref.~\cite{Degrande:2015vpa}, where fully-differential results in the 4FS were presented for the first time using
{\tt MG5\_aMC@NLO}~\cite{Alwall:2014hca} together with {\tt Herwig++}~\cite{Bahr:2008pv}
or {\tt Pythia8}~\cite{Sjostrand:2007gs}. 
The presence of a NLO+PS fully differential calculation allows a detailed comparison to the one in the 5FS.
It is interesting to observe that in Ref.~\cite{Degrande:2015vpa} it is shown that a reduced shower scale (by a factor of four) with respect to the default one in {\tt MG5\_aMC@NLO} improves the matching between parton shower and fixed-order results at large transverse momenta. Moreover the reduction of the matching scale choice also improves the agreement between the 4FS and 5FS calculations. It is interesting to notice that also in this case, as in the case of the choice of scales in the total cross section, a softer scale with respect to the naive hard scale is physically motivated and its employment improves the comparison between results in the two schemes. The inclusion of NLO(+PS) corrections further improves their mutual agreement at the level of shapes. 
Details of the comparison between schemes are illustrated in Fig.~\ref{fig:diffheavy} for two representative observables. On the left panel the transverse momentum distribution of the charged Higgs boson is plotted. In this case, as for all inclusive observables in the kinematics of the final bottom quark, the shapes of the four- and five-flavor scheme predictions agree very well. The difference in normalisation is compatible with the one observed in Ref.~\cite{Flechl:2014wfa}, although in this case it is more significant due to a slightly different choice of scale in the 5FS. Differences remain, however, and they are particularly sizeable for observables related to $b$ jets and $B$ hadrons, as it is displayed in the right panel of Fig.~\ref{fig:diffheavy}, in which the transverse momentum of the second-hardest $B$ hadron is plotted in the two schemes: at small $p_T$ the 4FS prediction is suppressed with respect to the 5FS one. This is due to mass effects: 
the $b$ quark is collinear to the beam. Such configurations are enhanced in a 5FS because of
the collinear singularities, while in the 4FS such a singularities are screened by the $b$-quark mass.
Given these differences, it is important to assess which are the most reliable predictions for this class of observables, given that the proper simulation of the signal is crucial to fully exploit the potential of the data collected in charged Higgs searches at the LHC. 
The recommendation of the authors of ~\cite{Degrande:2015vpa}, reiterated in the YR4~\cite{deFlorian:2016spz} is that 4FS predictions should be adopted for any realistic
signal simulation in experimental searches. Such recommendation is motivated by the fact that the 4FS prediction provides a better description of the final state kinematics and that the dependence on PS is smaller for the 4FS than for 5FS predictions. This is probably due to the fact that the 4FS has more differential information at the matrix-element level, which reduces the effects of the shower. For the normalisation of the cross section one could use the Santander-matched predictions of Ref.~\cite{Flechl:2014wfa}.

\section{Intermediate charged Higgs boson production}
\label{sec:intermediate}
\begin{figure}
\begin{center}
     \includegraphics[width=.4\textwidth]{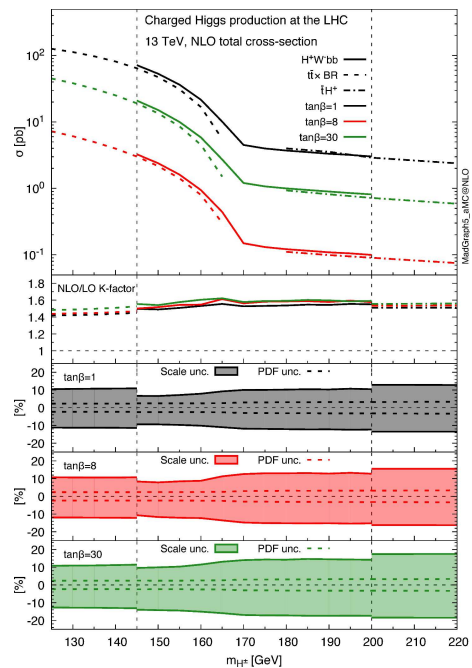}
\end{center}
\caption{\label{fig:intermediate} NLO total cross sections, $K$-factors and uncertainties for charged Higgs boson production at the 13 TeV LHC. The input parameters have been chosen consistently across all the mass range, in particular all cross sections are computed in the 4FS, the central scale for low-mass range is also set to $\mu=125$ GeV, while the scale $\mu = (m_t + m_{H^\pm} + m_b)/3$ is used for the heavy charged Higgs case.}
\end{figure}

The region in which the mass of the charged Higgs is close to the mass of
the top quark has not been explored so far by the experiments at the
LHC.  The main reason for that is the lack of precise theoretical
predictions for the charged Higgs production in that specific mass region.
Indeed, the treatment of the intermediate region between resonant
top quark decays and the continuum contribution for large charged
Higgs masses has been an open problem for some time. This has been
recently tackled by the full NLO calculation in the four-flavor scheme
published in Ref.~\cite{Degrande:2016hyf}. The calculation, performed in the complex-mass
scheme, for the intermediate top quarks allows to fully include
double- and single-resonant top contributions. Previous calculations
performed in several schemes were either done at leading order~\cite{Moretti:2002eu,Alwall:2004xw}
or by combining two processes without including
the full interference contributions between the two~\cite{Assamagan:2004gv,Weydert:2009vr,Klasen:2012wq}. 
As it is shown in Fig.~\ref{fig:intermediate}, the NLO QCD corrections computed in Ref.~\cite{Degrande:2016hyf} turn out to be large in this mass regime, with $K$-factors about 1.5-1.6. The central prediction in the main frame develop a prominent structure with a kink at the threshold $m_{H^\pm}\simeq m_t-m_b$.
Results nicely interpolate between the low- and high-mass regimes.
The effect of the single-resonant contributions ($pp\to t W^-$ and $pp \to \bar t H^+$) is visible when comparing 
the intermediate-mass result with the low-mass prediction. Indeed, 
the single-resonant contributions 
are missing in the low-mass prediction and amount to $10\%-15\%$ of the $pp \to t\bar t$ cross section depending on the specific
value of $\tan\beta$. Finally, looking at the matching of the intermediate-mass predictions to the heavy charged Higgs cross section, a $5\%-10\%$ gap can be observed for $\tan\beta=8$ and $\tan\beta=30$, originating from the non-resonant part of the $\hwbb$ amplitude, which, because of the chiral structure of the $H^+ tb$ and $Wtb$ vertices, is enhanced (suppressed) for large (small) values of $\tan \beta$. 

\section{Conclusions}
Important progress has been made in the past few years in the simulation of heavy and intermediate mass charged Higgs boson events, also thanks to the joint work of experimentalists and theorists in the Higgs Cross Section Working Group.
For heavy charged Higgs boson, Santander-matched predictions for wide range of masses and $\tan\beta$ in type-II 2HDM, generalisable to other types, have been made available for experimental searches. 
Furthermore fully differential calculations at NLO and NLO+PS are available both in the 5FS and in the 4FS. 
The comparison between 4FS versus 5FS comparison at the level of total and differential cross sections show that compatible results can be achieved for observables inclusive in bottom quark kinematics, also thanks to the choice of a lower shower and factorisation scales. Finally, the novel computation of total cross section calculation for simulation of signal in intermediate mass region is the first step towards an improved simulation of the differential distributions in that specific range, for which Run-II results will be soon made available.

\end{document}